\documentclass[twocolumn,unsortedaddress,superscriptaddress,showpacs,preprintnumbers]{revtex4}
\usepackage{graphicx}
\usepackage{epsfig}

\begin{document}

\title{Scattering of $^{7}$Be and $^{8}$B and the
astrophysical S$_{17}$ factor}

\affiliation{Cyclotron Institute, Texas A\&M University,College Station, TX 77843}
\affiliation{Institute of Nuclear Physics, Czech Academy of Science, Prague-Rez, Czech Republic}
\affiliation{Institute of Physics and Nuclear Engineering H. Hulubei, Bucharest, Romania}
\affiliation{}

\author{G.~Tabacaru}
\affiliation{Cyclotron Institute, Texas A\&M University,College Station, TX 77843}

\author{A.~Azhari}
\altaffiliation[Present address:  ]{Alix Partners, Dallas, TX}
\affiliation{Cyclotron Institute, Texas A\&M University,College Station, TX 77843}

\author{J.~Brinkley}
\affiliation{Cyclotron Institute, Texas A\&M University,College Station, TX 77843}

\author{V.~Burjan}
\affiliation{Institute of Nuclear Physics, Czech Academy of Science, Prague-Rez, Czech Republic}

\author{F.~Carstoiu}
\affiliation{Institute of Physics and Nuclear Engineering H. Hulubei, Bucharest, Romania}

\author{Changbo~Fu}
\affiliation{Cyclotron Institute, Texas A\&M University,College Station, TX 77843}

\author{C.A.~Gagliardi}
\affiliation{Cyclotron Institute, Texas A\&M University,College Station, TX 77843}

\author{V.~Kroha}
\affiliation{Institute of Nuclear Physics, Czech Academy of Science, Prague-Rez, Czech Republic}

\author{A.M.~Mukhamedzhanov}
\affiliation{Cyclotron Institute, Texas A\&M University,College Station, TX 77843}

\author{X.~Tang}
\altaffiliation[Present address:  ]{Argonne National Laboratory, Argonne, IL}
\affiliation{Cyclotron Institute, Texas A\&M University,College Station, TX 77843}

\author{L.~Trache}
\affiliation{Cyclotron Institute, Texas A\&M University,College Station, TX 77843}

\author{R.E.~Tribble}
\affiliation{Cyclotron Institute, Texas A\&M University,College Station, TX 77843}

\author{S. Zhou}
\affiliation{China Institute of Atomic Energy, P.O. Box 275(46), Beijing 102 413, P.R. China}

\date{\today }

\begin{abstract}
Measurements of scattering of $^{7}$Be at 87 MeV on a melamine (C$_{3}$N$%
_{6}$H$_{6}$) target and of $^{8}$B at 95 MeV on C were performed. For $^{7}$Be 
the angular range was extended over previous measurements and 
monitoring of the intensity of the radioactive beam was improved. The measurements allowed us to check and improve the optical model potentials used in the 
incoming and outgoing channels for the analysis of
existing data on the proton transfer reaction 
$^{14}$N($^{7}$Be,$^{8}$B)$^{13}$C.
The results
lead to an updated determination of the asymptotic
normalization coefficient for the virtual decay $^{8}$B $\rightarrow$ $^{7}$Be + $p$. 
We find a slightly larger value, $C_{tot}^{2}(^{8}B)=0.466\pm 0.047$ fm$^{-1}$, for the
melamine target.  This implies an astrophysical factor, $S_{17}(0)=18.0\pm 1.8$ eV$\cdot$b, 
for the solar neutrino generating reaction $^{7}$Be($p$,$\gamma $)$^{8}$B.
\end{abstract}
\pacs{24.10.Ht; 25.60.Bx; 25.60.Je; 26.65.+t} 
\maketitle


\section{Introduction}

Measurements of the energetic neutrinos produced in $^8$B beta decay
 have played a prominent role in our new understanding of neutrino properties
(see \cite{davis-prl20,fukuda-prl86,ahmed-prl87,bachall-prl92,haxton-ar05} and
references therein).
$^{8}$B is produced in the sun by the $^{7}$Be($p$,$\gamma $)$^{8}$B
reaction. A good understanding of this reaction rate is needed in order to calculate the expected
neutrino flux in the standard solar model \cite{bachall-prl92}. 
The determination of the astrophysical S$_{17}$ factor has,
therefore, been the subject of intense experimental and theoretical
effort over the past decade.
This work has been summarized in several recent publications 
\cite{davids-prc68,junghans-prc68,cyburt-prc70}. In spite of these efforts,
there is no clear consensus on the value of S$_{17}$(0) at the desired 5\%
precision. 
Consequently, several new experiments are under way or planned.

We previously reported measurements of the asymptotic normalization coefficients (ANC)
for $^8$B using the proton transfer reactions $^{10}$B($^{7}$Be,$%
^{8}$B)$^{9}$Be \cite{azhari-prl82} and $^{14}$N($^{7}$Be,$^{8}$B)$^{13}$C 
\cite{azhari-prc60}.  The ANCs
determine the amplitude of the tail
of the overlap integral of the ground state wave function of $^{8}$B onto
the two-body channel $^{7}$Be + $p$. 
To find the ANCs with the 
$^{14}$N($^{7}$Be,$^{8}$B)$^{13}$C transfer reaction, a distorted
wave Born approximation (DWBA) calculation is carried out and compared to the measured
data.  The DWBA calculation needs optical model
parameters (OMP) for both the incoming $^{7}$Be+$^{14}$N 
and outgoing $^{8}$B+$^{13}$C channels. Here we report
a measurement of $^{7}$Be elastic scattering on a melamine 
(C$_{3} $N$_{6}$H$_{6}$) target, where we doubled the angular range
and improved the monitoring of the intensity of the $^{7}$Be radioactive
beam relative to our previous measurement \cite{azhari-prc60}. 
The extension of the angular range was done to obtain a
better determination of the optical potential in the incoming channel. 
We also measured the elastic scattering of a $^{8}$B beam on a C target 
with the aim of checking, for the first
time, the OMP that were used in the outgoing channel $^{8}$B+$^{13}$C.

Below we describe the radioactive beam production, the experimental setups,
and the procedure for the data reduction. We then give results of calculations for optical
potentials in both the incoming and outgoing channels based on a double
folding procedure with an effective nucleon-nucleon interaction.  We discuss the
consequences of the improved secondary beam normalization, and 
compare revised DWBA calculations to the $^{14}$N($^{7}$Be,$^{8}$B)$^{13}$C data from Ref.\@ \cite{azhari-prc60} in order to
extract a new value for the ANCs and the corresponding astrophysical factor S$_{17}$.

\section{The experiments}

\begin{figure*}[t]
\includegraphics[height=8.5cm,width=15.2cm]{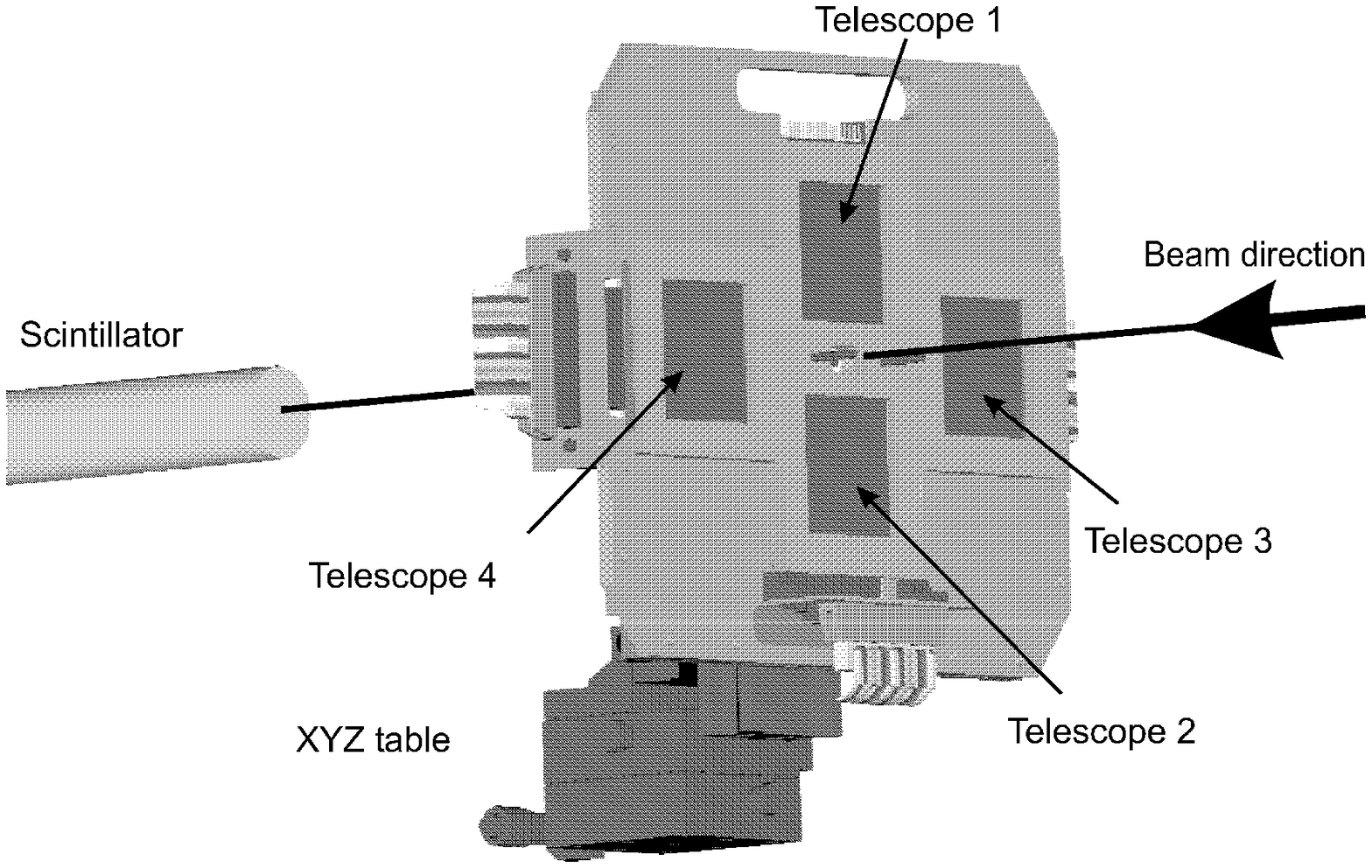} 
\caption{A three-dimensional view of the detector assembly.}
\label{view_3d}
\end{figure*}

The $^{7}$Be radioactive beam was produced and separated using the Momentum
Achromatic Recoil Spectrometer (MARS) \cite{tribble-npa701}. The primary beam was
$^{7}$Li at 18.6 MeV/A delivered by
the K500 superconducting cyclotron at Texas A\&M University.
It bombarded a
liquid nitrogen cooled H$_{2}$ gas target at a pressure of 2 atm, with entrance
and exit windows of 12 $\mu $m thick Havar. A secondary beam of $^{7}$Be at 12.5 MeV/A
was filtered from other reaction products by MARS. The characteristics of the beam
spot, which  were measured with a 900 $\mu $m thick two-dimensional Position Sensitive
Silicon Detector (PSSD) placed at the MARS focal plane (the target
detector), were a spot size of 2.5 mm $\times $ 3.6 mm FWHM
(horizontal $\times $ vertical) and an angular spread of $1.8^{\circ }\times
0.6^{\circ }$. The purity of the $^{7}$Be beam was 99\% at an average rate of 
$\sim$\,80 kHz.  Alpha particles were the primary contaminant. For a detailed description of radioactive
beam production with MARS, see Ref.\@ \cite{azhari-prc63}. Following beam tuning, the
secondary target, a melamine foil with a thickness of 1.5
mg/cm$^{2}$, was moved into the beam spot. 
Four $5\times 5$ cm$^{2}$ PSSDs were placed
symmetrically around the target on an aluminum plate, as shown in Fig.\@ 
\ref{view_3d}. Detectors 1 and 2 (110 $\mu $m thick) covered a laboratory angular
range from 4${^{\circ}}$ to 19${^{\circ}}$, and detectors 3 and 4 (65 $\mu $m thick) covered 
16${^{\circ}}$  to 30${^{\circ}}$.
All four PSSDs were backed by 500 $\mu$m thick silicon detectors providing particle
identification spectra ($\Delta E,E$). Each PSSD was position calibrated using a mask with 6
slots that were 0.8~mm wide and were spaced  8 mm apart. The detectors were cooled to
approximately $-10^{\circ }$C with two
electric thermocoolers fixed on the aluminum plate in order to decrease the
inverse current in the detectors and minimize their noise. The assembly was
placed on a XYZ optical table for precise positioning.

In our previous experiments with a $^{7}$Be beam \cite{azhari-prl82,azhari-prc60}, 
the number of secondary beam particles
was determined indirectly by measuring the intensity of the $^7$Li primary beam in
a Faraday cup, and normalizing the yield at low primary beam intensities by
counting the $^{7}$Be with the target detector.  Periodically (typically once a day) the
calibration 
procedure was repeated to check for any rate variations due to drifts in MARS power
supplies.  The primary $^{7}$Li beam intensity was substantially higher for the 
experiment on the melamine target \cite{azhari-prc60} than the experiment on the $^{10}$B
\cite{azhari-prl82} 
target.

Following these two measurements, we modified 
the experimental setup by adding a monitor detector to count the radioactive beam particles
directly. The
beam monitoring system, which is shown in Fig.\@ \ref{view_3d}, used a wire mesh screen to
reduce the secondary beam intensity and a plastic scintillator
coupled with a photomultiplier tube to count the radioactive beam particles that passed through
the target. In parallel, we ran the old monitoring system with a
Faraday cup for the primary beam, and compared the results. 
In subsequent measurements with a high-intensity $^{11}$B primary beam, we observed a
difference between the two normalization procedures.  Beam heating reduced the density of the gas in the production target, causing a drop in the isotope production rate per nA of
primary beam current and increasing the central beam energy. 

By scaling the heat deposition of the beam in the gas target, we concluded that this effect may have produced a small
but non-negligible shift in the beam normalization during the 
previous $^{14}$N($^{7}$Be,$^8$B)$^{13}$C experiment.  The effect was to over estimate the number of
secondary beam particles and hence reduce the cross section and the resulting ANC.  
In contrast, the primary beam intensity for the $^{10}$B($^7$Be,$^8$B)$^9$Be measurement was sufficiently low to have a negligible effect.
For the present experiment, the monitor detector was a NE102A plastic
scintillator coupled to a photomultiplier tube.  To minimize rate-dependent effects in
the photomultiplier tube, two screens with a transparency of 9\% each were
added to attenuate the beam intensity.  The yield in the monitor detector was calibrated using the
procedure described in Ref.\@ \cite{tang-prc67}. 

For the $^{8}$B elastic scattering measurement, the radioactive beam was produced via the
$^{1}$H($^{10}$B,$^{8}$B)$^{3}$H reaction using a 27 MeV/A $^{10}$B
primary beam on the same LN$_{2}$-cooled gas cell. The cell contained H$_{2}$
gas at 3 atm pressure, corresponding to a target thickness of $\approx $\thinspace
10.8 mg/cm$^{2}$. Entrance and exit windows were made of 50 $\mu $m (42 mg/cm$^{2})$
Havar. A 137 mg/cm$^{2}$ Al degrader was
placed behind the gas cell to reduce the secondary beam energy. A
95 MeV radioactive beam of $^{8}$B was focused at the end of MARS with a rate of about 
5 kHz. The beam purity was better than 95\%, with $\alpha $ particles being the
primary contaminant.  The full-width energy spread was limited to 1.6 MeV using 
momentum defining slits. Beam emittance was
optimized using a pair of slits after the last quadrupole in MARS.  
The plastic scintillator behind the target was used for direct counting of the
secondary beam particles, in this case without any wire mesh screen.  Two telescopes, each consisting of
a 110 $\mu$m thick PSSD backed by a 500 $\mu$m thick Si detector, observed the secondary reaction products.  The telescopes covered the
angular range $\theta _{lab}=4^{\circ }-19^{\circ }$.
A 1.9 mg/cm$%
^{2}$ C target was used for the elastic scattering measurement.

In both experiments,
target properties such as
thickness and uniformity were verified using the radioactive $^{7}$Be and 
$^{8}$B beams directly by detecting beam
particles at $0^{\circ}$ with and without the target. The resultant energy
loss measurements were compared to calculations with
the computer code SRIM \cite{Srim99} to extract the thicknesses.

\section{\label{sec_data}Results}

\subsection{$^{7}$Be elastic scattering}

There were two motivations for measuring elastic scattering of $^{7}$Be from melamine.  The
new detector geometry allowed us to extend the angular region for elastics which, in turn, helps
to define optical model parameters.  Further, it allowed us to normalize the elastic
scattering yield directly by counting $^{7}$Be
particles after the secondary target.  

\begin{figure}[tbp]
\includegraphics[width=8.5cm,height=8.5cm]{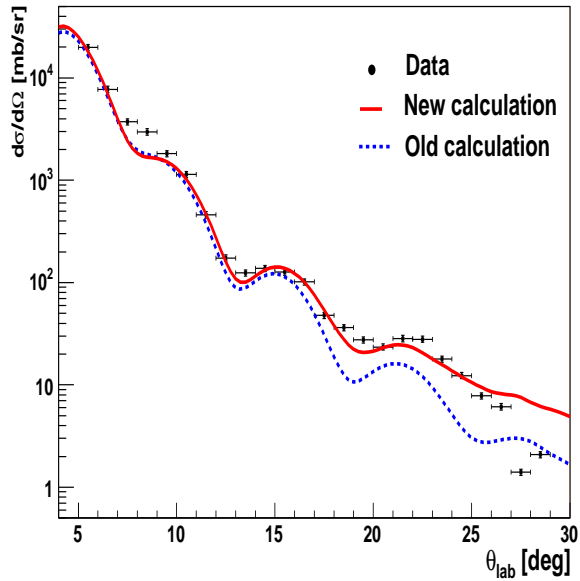}
\caption{(color online) Angular distribution for elastic scattering of $^{7}$%
Be on $^{14}$N and $^{12}$C. The points are experimental data after subtraction
of the inelastic scattering contribution. The solid (red) line is the new calculation, and
the dashed line (blue) is the previous calculation from Ref.\@ \protect\cite{azhari-prc63}. Both
calculations are smeared to account for the finite angular resolution.}
\label{fig_ang_dis}
\end{figure}

The kinematic reconstruction of the elastically scattered  reaction products was
performed using the energy and position information from the
four detector telescopes. The events selected corresponded
to elastic scattering of $^{7}$Be on $^{14}$N and $^{12}$C since the two
contributions could not be separated. 
First, we identified all the events with ($\Delta E,E$) corresponding to $^{7}$Be,
then we utilized the correlation of $^7$Be energy vs.\@ scattering angle to select those that were consistent with elastic scattering off either $^{14}$N or $^{12}$C.  This discriminated against scattering on H and inelastic scattering populating excited states in either $^{14}$N or $^{12}$C.  However, it was not possible to separate the elastically scattered
events from inelastic scattering
leading to the first excited state
in $^{7}$Be at $E_{ex}$ = 0.429 MeV.  We estimated this
contribution using data
obtained from inelastic scattering of $^{7}$Li on $^{13}$C at 63 MeV \cite{trache-prc67} to
the analog state at $E_{ex}$ = 0.477
MeV.  Assuming that the
deformation lengths are equal for the analog states, we calculated the inelastic cross
section for $^{7}$Be on $^{14}$N and $^{12}$C at the current energy and subtracted it from
the data. The correction 
was negligible at all but the largest angles, where it amounted to a
few percent.
The solid angle calculation as a function of scattering angle was done
using a Monte Carlo simulation that took the measured properties of the beam spot and the
geometrical specifications of the detector assembly as input data. The procedure has been described in previous
publications (e.g. see Ref.\@ \cite{azhari-prc63}).  Figure \ref{fig_ang_dis} shows the resulting angular distribution corresponding to elastically scattered $^{7}$Be on $^{14}$N and $^{12}$C.

The angular distribution predicted from the optical model parameters used in
Ref.\@ \cite{azhari-prc63} is compared to the data in Fig.\@ \ref{fig_ang_dis}. The Monte Carlo
calculation was used to provide the proper angular distribution that takes into account the finite
angular binning in the data.  

The results of our measurements are compatible with those reported in Ref.\@ \cite{azhari-prc60}
at small angles and with the predictions of the optical model calculations
done at that time (dotted curve). However, the new experimental data fall above the predictions
at larger angles, suggesting a smaller absorption than was
assumed in Ref.\@ \cite{azhari-prc60}.  In order to obtain a better description of the elastic scattering, 
calculations were carried out with a range of new parameters.  These were also used for the
entrance channel to generate DWBA predictions for the
$^{14}$N($^{7}$Be,$^{8}$B)$^{13}$C proton transfer
reaction.

The optical parameters used in Ref.\@ \cite{azhari-prc60} were based on results from an analysis of
elastic scattering of loosely bound 
$p$-shell nuclei \cite{trache-prc61}, which demonstrated that the data can be described
with double-folded potentials. The potentials quoted in Ref.\@ \cite{trache-prc61} were
obtained from calculated nuclear matter densities folded with an 
effective nucleon-nucleon interaction (JLM, \cite{JLM}), smeared (two range
parameters, $t_{V}$ and $t_{W}$) and renormalized (two strength parameters, 
$N_{V}$ and $N_{W}$) to produce: 
\begin{equation}
U_{DF}(r)=N_{V}V(r,t_{V})+iN_{W}W(r,t_{W}) .
\label{fp1}
\end{equation}%
The calculations for previous $^{7}$Be studies  were done using the JLM1 effective interaction
with standard
range parameters: $t_{V}$=1.20 fm, $t_{W}$=1.75 fm, and average
renormalizations $N_{V}$=0.37, $N_{W}$=1.00 (for details see Ref.\@ \cite{trache-prc61} and
references therein). 
These parameters served as the starting point for the new calculations.  
Elastic scattering of $^{7}$Be at 87.7 MeV on $^{14}$N
and $^{12}$C were calculated in the center of mass frame, then 
transformed into the lab frame and added with weights 1.0 and 0.5,
respectively, equal to the ratio of $^{14}$N to $^{12}$C nuclei in the melamine. The resulting
curve was ``smoothed" using the Monte Carlo
code described above. The parameters for the
folding potential were varied simultaneously and identically for both target
nuclei. This approach is supported by the fact that
both target nuclei are well bound and have similar densities in the surface
region and by experiments we have carried out with melamine and C targets using other
radioactive beams, such as $^{13}$N \cite{tang-prc69}, $^{8}$B (present
experiment) and $^{17}$F \cite{blackmon-npa746}. The extended angular coverage of the
present data was still not sufficient to attempt an optical model fit with free parameters.
Rather, the two normalization and two range parameters  were varied. 
The parameters for various calculations and the reduced $\chi ^{2}$ values
obtained by comparison to the data are presented in Table \ref{tab_par}.

Four entries in the Table (A, B, C and H) were obtained by adjusting the 
renormalization of the real and imaginary parts of the potential.
The smearing ranges of the interaction, $t_{V}$ and 
$t_{W}$, were adjusted for three cases (D, E, F), and the density dependence was
adjusted in one case (G) where the JLM2 interaction was used.
The best results were obtained for cases D, E, G and H. The small differences
between the $\chi{^2}$ values show that it is difficult to choose a
``best solution". Rather, we did DWBA
calculations for the $^{14}$N($^7$Be,$^8$B)$^{13}$C transfer reaction for the four most promising potentials.  The results are compared to the
previous calculations in Table \ref{tab_par}.
 
\begin{figure}[tbp]
\includegraphics[width=9.cm,height=12cm]{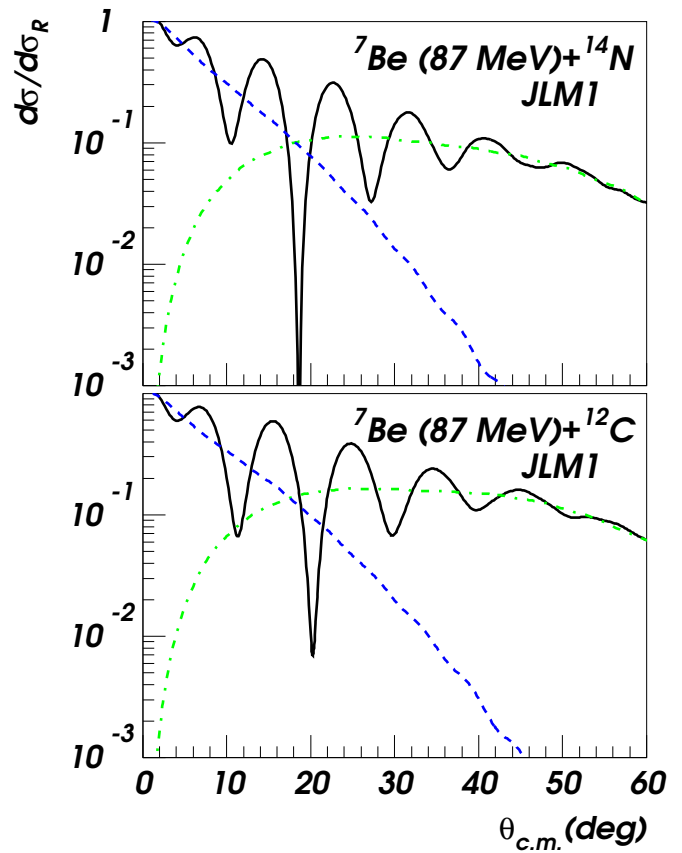}
\caption{(color online) Far side/near side components shown compared as ratios to Rutherford
scattering for the potential H in Table \protect\ref{tab_par}.  The near side components are the 
dashed lines, the far side components are the dot-dashed lines, and the totals are given as the solid lines.}
\label{fig_FN}
\end{figure}

A far/near decomposition of the scattering amplitudes shows that the observed angular range 
covers the region of Fraunhofer oscillations generated by the
interference of the two components (see Fig.\@ \ref{fig_FN}). Their
crossover is around 20$^\circ$, and at larger angles the far component
becomes dominant.  But in the region included in our measurements, the
interference is still important. The calculations show that after about 60$^{\circ}$
the angular distributions develop a rainbow type pattern, typical
for the cases found recently in our $^{6,7}$Li elastic scattering data \cite{carstoiu-prc70}.

\subsection{$^{8}$B experiment}

\begin{figure}[tbp]
\includegraphics[width=9.cm,height=9cm]{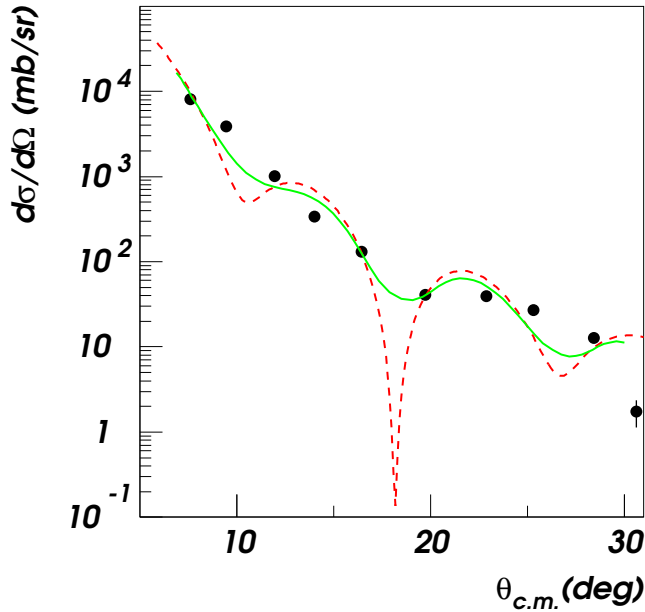}
\caption{(color online) The angular distribution for elastic scattering
of $^8$B on C. The dashed line is the calculated cross section and the solid line is the result after 
accouting for the finite angular resolution.}
\label{ang-dist-8B}
\end{figure}

A similar analysis was used for the $^{8}$B elastic scattering on a natural C target. The 
resulting angular distribution is shown in Fig.\@ \ref{ang-dist-8B} where it is compared to
calculations made with the folded potentials using the average parameters $t_{V}=1.20$ fm,
$t_{W}=1.75$ fm, $N_{V}=0.37$ and $N_{W}=1.00$. The solid (dashed) line shows the
results after (before) smoothing with the Monte Carlo calculation.  The
$^{8}$B density used in the folding procedure was that
calculated in \cite{trache-prc61} using the correct ANC for the last proton.
Due to the limited angular range of the data, we did not attempt to produce a better fit by
adjusting parameters.
Based on the similar
densities for $^{12}$C and $^{13}$C and on results found in cases where 
scattering on both $^{12}$C and $^{13}$C were measured, we assume that the
parameters ($t_{V}$, $t_{W} $, $N_{V}$, $N_{W}$) extracted for the natural C target are
valid for the $^{8}$B+$^{13}$C channel in DWBA calculations of the transfer reaction.

\section{Implications for $^{14}$N($^7$Be,$^8$B)$^{13}$C}

Based on the present analysis, the 
value of the $^{14}$N($^7$Be,$^8$B)$^{13}$C proton transfer reaction cross section from our previous measurement (\cite{azhari-prc60})
was increased,
$\sigma_{exp}^{tr}(new)=1.055\,\sigma _{exp}^{tr}(old)$, to account for the difference of
5.5\% found in the absolute normalization. Following the original publication the ratio between
the ANCs for the $1p_{1/2}$ and $1p_{3/2}$ components in the wave function of
the ground state of $^{8}$B was found to be
$\delta ^{2}=C^{2}(^{8}B,p_{1/2})/$ $C^{2}(^{8}B,p_{3/2})=$ $0.125(20)$ 
from the mirror neutron transfer
reaction $^{13}$C($^{7}$Li,$^{8}$Li)$^{12}$C \cite{trache-prc67}. The 
previous result had assumed $\delta ^{2}=0.157$ based on theoretical calculations. The new
cross section, the value of $\delta ^{2}$ from experiment, and the new optical model parameters have
been used to find revised values of the ANCs for $^8$B $\to$ $^7$Be + $p$ from the $^{14}$N($^7$Be,$^8$B)$^{13}$C proton transfer reaction measurement \cite{azhari-prc60}.  

In order to extract the ANCs, four terms must be calculated to account for 
the possible proton transitions:
\begin{widetext}
\begin{eqnarray}
\frac{d\sigma }{d\Omega } &=&C_{^{8}B,p_{3/2}}^{2}\left[ \frac{%
C_{^{14}N,p_{1/2}}^{2}}{b_{^{14}N,p_{1/2}}^{2}}\frac{\sigma
_{p_{1/2}\rightarrow p_{3/2}}}{b_{^{8}B,p_{3/2}}^{2}}+\frac{%
C_{^{14}N,p_{3/2}}^{2}}{b_{^{14}N,p_{3/2}}^{2}}\frac{\sigma
_{p_{3/2}\rightarrow p_{3/2}}}{b_{^{8}B,p_{3/2}}^{2}}\right]   \nonumber \\
&+&C_{^{8}B,p_{1/2}}^{2}\left[ \frac{C_{^{14}N,p_{1/2}}^{2}}{%
b_{^{14}N,p_{1/2}}^{2}}\frac{\sigma _{p_{1/2}\rightarrow p_{1/2}}}{%
b_{^{8}B,p_{1/2}}^{2}}+\frac{C_{^{14}N,p_{3/2}}^{2}}{b_{^{14}N,p_{3/2}}^{2}}%
\frac{\sigma _{p_{3/2}\rightarrow p_{1/2}}}{b_{^{8}B,p_{1/2}}^{2}}\right]
=C^{2}(^{8}B,p_{3/2})\left[ \widetilde{\sigma }_{DW}\right]   \label{eq_anc}
\end{eqnarray}
\end{widetext}
where the $\sigma $'s are the calculated DWBA differential
cross sections for proton transfer from the $p_{3/2}$ and $p_{1/2}$ orbitals
in $^{14}$N to the $p_{3/2}$ and $p_{1/2}$ orbitals in $^{8}$B, the $b_{lj}$'s
are the ANCs for the single particle orbitals
used in the DWBA calculation, and the $C_{^{14}N,lj}$'s and $C_{^{8}B,lj}$'s are
the ANCs for ${^{14}}$N $\rightarrow$ ${^{13}}$C + $p$ and 
${^{8}}$B $\rightarrow$ $^{7}$Be + $p$, respectively. 
The ANCs, $C_{^{14}N,p_{1/2}}^{2}$=18.6(12)~fm$^{-1}$ and 
$C_{^{14}N,p_{3/2}}^{2}$=0.93(14)~fm$^{-1}$, were measured in
\cite{trache-prc58,bem-prc62}. The calculations were done with
the code PTOLEMY \cite{PTOLEMY}. The
results of the calculations are given in Table \ref{tab_par} where the value calculated and
shown in column 7 is the quantity in the last square bracket in Eq.\@ (\ref{eq_anc}),
$\widetilde{\sigma }_{DW}$, integrated over the 
angular region $\theta_{c.m.}$ = 4${{}^{\circ }}$- 25${{}^{\circ }}$ to match the data. 
This quantity contains the DWBA cross
sections weighted with the ANCs for $^{14}$N, the
single particle ANCs calculated for the appropriate Woods-Saxon proton
binding potentials in $^{8}$B and $^{14}$N and the mixing ratio in the
ground state of $^{8}$B, $\delta ^{2}$. 
Since the reaction is peripheral, the results do not depend on the
geometry assumed for the proton binding potentials, which are chosen to be
Woods-Saxon shape with depths adjusted to reproduce the experimental proton
binding energies in $^{8}$B and $^{14}$N, respectively. The results shown
were calculated using the reduced radius $r_{0}=1.20$ fm and the diffuseness 
$a=0.60$ fm, and the same spin-orbit term as in Ref.\@ \cite{azhari-prc60}. 
The exit channel parameters were fixed to the previous values.
Calculations were done at $E_{lab}=83.5$ MeV, the energy of the previous
experiment with the four optical model sets in Table \ref{tab_par} for the entrance channel that
have the lowest 
$\chi{^{2}}$, B, D, E, G, and H.  In column 8 we give the
ratio of the present calculations to the same quantity calculated in Ref.\@ \cite{azhari-prc60}.
The average of the four results, weighted by the $\chi{^{2}}$'s, gives the ratio
$\left\langle \mathcal{R}\right\rangle =0.968\pm 0.047$.
The new ANC, calculated with the relation
\begin{equation}
{C_{^{8}B,p_{3/2}}^{2}(new)=(1.055/\left\langle \mathcal{R}\right\rangle
)C_{^{8}B,p_{3/2}}^{2}(old),}  
\label{old-new}
\end{equation}
is $C_{^{8}B,p_{3/2}}^{2}(new)=0.414\pm 0.041$ fm$%
^{-1}$.  The
overall uncertainty contains contributions from statistics (2.6\%), absolute
normalization of the cross section (5\%), input parameters in the Monte
Carlo simulation of the experiment (1.4\%), and uncertainties in the ANC for
the $^{14}$N vertex (6.4\%). The contribution of each of these factors
remains the same as in the original analysis \cite{azhari-prc60}. The
uncertainty due to the optical model parameters was taken from the standard
deviation of the calculated cross sections (column 7 or 8 in Table \ref{tab_par}) 
and is 5\% compared with the previous value of 8.1\%.

The relation between the ANCs and the astrophysical factor $S_{17}$(0), in eV$\cdot$b,
is \cite{hmxu-prl94}
\begin{equation}
S_{17}(0)=38.6\left( C_{p_{3/2}}^{2}+C_{p_{1/2}}^{2}\right)
=38.6C_{p_{3/2}}^{2}(1+\delta ^{2})
\label{eq_s17}
\end{equation}%
Using the new value of the ANC we find $S_{17}(0)=18.0\pm 1.8$ eV$\cdot$b. 
This value is very close to the value obtained from
the reaction $^{10}$B($^{7}$Be,$^{8}$B)$^{9}$Be in Ref.\@ \cite{azhari-prl82} 
where we found $S_{17}(0)=18.4\pm 2.5$ eV$\cdot$b. The weighted average of the two results
is $S_{17}(0)=18.2\pm 1.7$ eV$\cdot$b.

\section{Conclusion}

Elastic scattering of $^{7}$Be at about 12 MeV/A has been measured over an 
extended angular range on a melamine target. 
The results provide a better determination of the optical
model parameters used for the entrance channel of the 
$^{14}$N($^{7}$Be,$^{8}$B)$^{13}$C reaction. For the
first time elastic scattering of $^{8}$B was measured on a C target, thus allowing for a check
the optical model parameters used for the exit channel in the DWBA calculation. In the
measurement of the $^{7}$Be elastic scattering, we directly counted the secondary beam
particles. This resulted in a 5.5\% increase of the transfer reaction cross section from Ref.\@
\cite{azhari-prc60}, which used an indirect method to obtain the secondary beam intensity. 
We also used the mixing ratio between the $1p_{1/2}$ and $1p_{3/2}$ components from a
($^{7}$Li,$^{8}$Li) measurement \cite{trache-prc67}, rather than a theoretical prediction. 
These improvements lead to
the revised value of the ANC for $^{8}$B $\rightarrow$ ${^{7}}$Be + $p$ from
the $^{14}$N($^{7}$Be,$^{8}$B)$^{13}$C reaction of 
$C^{2}(^{8}B,p_{3/2};new)=0.414\pm 0.041$ fm$^{-1},$ resulting in 
$C_{tot}^{2}(^{8}B;new)=C_{p_{3/2}}^{2}+C_{p_{1/2}}^{2}=0.466\pm 0.047$ fm$%
^{-1}$.  This, in turn leads to a larger value for the astrophysical S factor for the $^{7}$%
Be(p,$\gamma $)$^{8}$B reaction, $S_{17}(0)=18.0\pm 1.8$ eV$\cdot$b. This
new value is very close to the one from the same reaction on the 
$^{10}$B target \cite{azhari-prl82}. Averaging the two results, we obtain 
$S_{17}(0)=18.2\pm 1.7$ eV$\cdot$b from the proton transfer reactions.

Our result for $S_{17}(0)$ is a bit over 2$\sigma$ lower than the extrapolation of the most
recent and precise direct measurement of ${^{7}}$Be(p,$\gamma$)${^{8}}$B by Junghans 
{\it et al.} \cite{junghans-prc68}.  Our central value is about 1.5$\sigma$ lower than the average
central value obtained by Cyburt {\it et al.} \cite{cyburt-prc70} in a recent analysis that uses all
of the best available capture data, under the assumption that they are independent. Including the
uncertainty quoted by Cyburt {\it et al.} our results are consistent at the 1$\sigma$ level.  
We do not understand the reason for the discrepancy between our ANC result and the
extrapolated value from Junghans {\it et al.}. However, we note that direct measurements with
both radioactive beams and targets and indirect measurements continue to be carried out on this
important proton capture reaction.  


\acknowledgments

This work was supported in part by the U.S. Department of Energy under Grant
No. DE-FG03-93ER40773, the U.S. National Science Foundation under Grant No.
INT-459521-00001, the Romanian Ministry for Education, Research and Youth under
contract no. 555/2000, and the Robert A. Welch Foundation. One of the authors
(F.C.) acknowledges the support of the Cyclotron Institute, Texas A\&M
University for part of the time this work was done.

\bibliography{be7n14}

\begin{thebibliography}{24}
\expandafter\ifx\csname natexlab\endcsname\relax\def\natexlab#1{#1}\fi
\expandafter\ifx\csname bibnamefont\endcsname\relax
  \def\bibnamefont#1{#1}\fi
\expandafter\ifx\csname bibfnamefont\endcsname\relax
  \def\bibfnamefont#1{#1}\fi
\expandafter\ifx\csname citenamefont\endcsname\relax
  \def\citenamefont#1{#1}\fi
\expandafter\ifx\csname url\endcsname\relax
  \def\url#1{\texttt{#1}}\fi
\expandafter\ifx\csname urlprefix\endcsname\relax\def\urlprefix{URL }\fi
\providecommand{\bibinfo}[2]{#2}
\providecommand{\eprint}[2][]{\url{#2}}

\bibitem[{\citenamefont{Davis et~al.}(1968)\citenamefont{Davis, Harmer, and
  Hoffman}}]{davis-prl20}
\bibinfo{author}{\bibfnamefont{R.}~\bibnamefont{Davis}},
  \bibinfo{author}{\bibfnamefont{D.}~\bibnamefont{Harmer}}, \bibnamefont{and}
  \bibinfo{author}{\bibfnamefont{K.~C.} \bibnamefont{Hoffman}},
  \bibinfo{journal}{Phys. Rev. Lett} \textbf{\bibinfo{volume}{20}},
  \bibinfo{pages}{1205} (\bibinfo{year}{1968}).

\bibitem[{\citenamefont{{S.~Fukuda~et~al.}}(2001)}]{fukuda-prl86}
\bibinfo{author}{\bibnamefont{{S.~Fukuda~et~al.}}}
  (\bibinfo{collaboration}{Super-Kamiokande collaboration}),
  \bibinfo{journal}{Phys. Rev. Lett} \textbf{\bibinfo{volume}{86}},
  \bibinfo{pages}{5651} (\bibinfo{year}{2001}).

\bibitem[{\citenamefont{{S.~N.~Ahmed et al.}}(2003)}]{ahmed-prl87}
\bibinfo{author}{\bibnamefont{{S.~N.~Ahmed et al.}}}
  (\bibinfo{collaboration}{{SNO collaboration}}), \bibinfo{journal}{Phys. Rev.
  Lett} \textbf{\bibinfo{volume}{87}}, \bibinfo{pages}{071301}
  (\bibinfo{year}{2003}).

\bibitem[{\citenamefont{Bahcall and Pinsonneault}(2004)}]{bachall-prl92}
\bibinfo{author}{\bibfnamefont{J.}~\bibnamefont{Bahcall}} \bibnamefont{and}
  \bibinfo{author}{\bibfnamefont{M.}~\bibnamefont{Pinsonneault}},
  \bibinfo{journal}{Phys. Rev. Lett} \textbf{\bibinfo{volume}{92}},
  \bibinfo{pages}{121301} (\bibinfo{year}{2004}).

\bibitem[{\citenamefont{Haxton et~al.}(2005)\citenamefont{Haxton, Parker, and
  Rolfs}}]{haxton-ar05}
\bibinfo{author}{\bibfnamefont{W.}~\bibnamefont{Haxton}},
  \bibinfo{author}{\bibfnamefont{P.}~\bibnamefont{Parker}}, \bibnamefont{and}
  \bibinfo{author}{\bibfnamefont{C.}~\bibnamefont{Rolfs}},
  \bibinfo{journal}{arxiv:nucl-th/0501020}  (\bibinfo{year}{2005}).

\bibitem[{\citenamefont{Davids and Typel}(2003)}]{davids-prc68}
\bibinfo{author}{\bibfnamefont{B.}~\bibnamefont{Davids}} \bibnamefont{and}
  \bibinfo{author}{\bibfnamefont{S.}~\bibnamefont{Typel}},
  \bibinfo{journal}{Phys. Rev. C} \textbf{\bibinfo{volume}{68}},
  \bibinfo{pages}{045802} (\bibinfo{year}{2003}).

\bibitem[{\citenamefont{Junghans et~al.}(2003)\citenamefont{Junghans, Mohrmann,
  Snover, Steiger, Adelberger, Casandjian, , Swanson, Buchmann, Park
  et~al.}}]{junghans-prc68}
\bibinfo{author}{\bibfnamefont{A.~R.} \bibnamefont{Junghans}},
  \bibinfo{author}{\bibfnamefont{E.~C.} \bibnamefont{Mohrmann}},
  \bibinfo{author}{\bibfnamefont{K.~A.} \bibnamefont{Snover}},
  \bibinfo{author}{\bibfnamefont{T.~D.} \bibnamefont{Steiger}},
  \bibinfo{author}{\bibfnamefont{E.~G.} \bibnamefont{Adelberger}},
  \bibinfo{author}{\bibfnamefont{J.~M.} \bibnamefont{Casandjian}}, ,
  \bibinfo{author}{\bibfnamefont{H.~E.} \bibnamefont{Swanson}},
  \bibinfo{author}{\bibfnamefont{L.}~\bibnamefont{Buchmann}},
  \bibinfo{author}{\bibfnamefont{S.~H.} \bibnamefont{Park}},
  \bibnamefont{et~al.}, \bibinfo{journal}{Phys. Rev. C}
  \textbf{\bibinfo{volume}{68}}, \bibinfo{pages}{065803}
  (\bibinfo{year}{2003}).

\bibitem[{\citenamefont{Cyburt et~al.}(2004)\citenamefont{Cyburt, Davids, and
  Jennings}}]{cyburt-prc70}
\bibinfo{author}{\bibfnamefont{R.~H.} \bibnamefont{Cyburt}},
  \bibinfo{author}{\bibfnamefont{B.}~\bibnamefont{Davids}}, \bibnamefont{and}
  \bibinfo{author}{\bibfnamefont{B.~K.} \bibnamefont{Jennings}},
  \bibinfo{journal}{Phys. Rev. C} \textbf{\bibinfo{volume}{70}},
  \bibinfo{pages}{045801} (\bibinfo{year}{2004}).

\bibitem[{\citenamefont{Azhari et~al.}(1999{\natexlab{a}})\citenamefont{Azhari,
  Burjan, Carstoiu, Dejbakhsh, Gagliardi, Kroha, Mukhamedzhanov, Trache, and
  Tribble}}]{azhari-prl82}
\bibinfo{author}{\bibfnamefont{A.}~\bibnamefont{Azhari}},
  \bibinfo{author}{\bibfnamefont{V.}~\bibnamefont{Burjan}},
  \bibinfo{author}{\bibfnamefont{F.}~\bibnamefont{Carstoiu}},
  \bibinfo{author}{\bibfnamefont{H.}~\bibnamefont{Dejbakhsh}},
  \bibinfo{author}{\bibfnamefont{C.}~\bibnamefont{Gagliardi}},
  \bibinfo{author}{\bibfnamefont{V.}~\bibnamefont{Kroha}},
  \bibinfo{author}{\bibfnamefont{A.}~\bibnamefont{Mukhamedzhanov}},
  \bibinfo{author}{\bibfnamefont{L.}~\bibnamefont{Trache}}, \bibnamefont{and}
  \bibinfo{author}{\bibfnamefont{R.}~\bibnamefont{Tribble}},
  \bibinfo{journal}{Phys. Rev. Lett.} \textbf{\bibinfo{volume}{82}},
  \bibinfo{pages}{3960} (\bibinfo{year}{1999}{\natexlab{a}}).

\bibitem[{\citenamefont{Azhari et~al.}(1999{\natexlab{b}})\citenamefont{Azhari,
  Burjan, Carstoiu, Gagliardi, Kroha, Mukhamedzhanov, Tang, Trache, and
  Tribble}}]{azhari-prc60}
\bibinfo{author}{\bibfnamefont{A.}~\bibnamefont{Azhari}},
  \bibinfo{author}{\bibfnamefont{V.}~\bibnamefont{Burjan}},
  \bibinfo{author}{\bibfnamefont{F.}~\bibnamefont{Carstoiu}},
  \bibinfo{author}{\bibfnamefont{C.~A.} \bibnamefont{Gagliardi}},
  \bibinfo{author}{\bibfnamefont{V.}~\bibnamefont{Kroha}},
  \bibinfo{author}{\bibfnamefont{A.~M.} \bibnamefont{Mukhamedzhanov}},
  \bibinfo{author}{\bibfnamefont{X.}~\bibnamefont{Tang}},
  \bibinfo{author}{\bibfnamefont{L.}~\bibnamefont{Trache}}, \bibnamefont{and}
  \bibinfo{author}{\bibfnamefont{R.~E.} \bibnamefont{Tribble}},
  \bibinfo{journal}{Phys. Rev. C} \textbf{\bibinfo{volume}{60}},
  \bibinfo{pages}{055803} (\bibinfo{year}{1999}{\natexlab{b}}).

\bibitem[{\citenamefont{Tribble et~al.}(2002)\citenamefont{Tribble, Azhari,
  Gagliardi, Hardy, Mukhamedzhanov, Tang, Trache, and
  Yennello}}]{tribble-npa701}
\bibinfo{author}{\bibfnamefont{R.~E.} \bibnamefont{Tribble}},
  \bibinfo{author}{\bibfnamefont{A.}~\bibnamefont{Azhari}},
  \bibinfo{author}{\bibfnamefont{C.~A.} \bibnamefont{Gagliardi}},
  \bibinfo{author}{\bibfnamefont{J.~C.} \bibnamefont{Hardy}},
  \bibinfo{author}{\bibfnamefont{A.~M.} \bibnamefont{Mukhamedzhanov}},
  \bibinfo{author}{\bibfnamefont{X.}~\bibnamefont{Tang}},
  \bibinfo{author}{\bibfnamefont{L.}~\bibnamefont{Trache}}, \bibnamefont{and}
  \bibinfo{author}{\bibfnamefont{S.~J.} \bibnamefont{Yennello}},
  \bibinfo{journal}{Nucl. Phys. A} \textbf{\bibinfo{volume}{701}},
  \bibinfo{pages}{278C} (\bibinfo{year}{2002}).

\bibitem[{\citenamefont{Azhari et~al.}(2001)\citenamefont{Azhari, Burjan,
  Gagliardi, Kroha, Mukhamedzanov, Nunes, Tang, Trache, and
  Tribble}}]{azhari-prc63}
\bibinfo{author}{\bibfnamefont{A.}~\bibnamefont{Azhari}},
  \bibinfo{author}{\bibfnamefont{V.}~\bibnamefont{Burjan}},
  \bibinfo{author}{\bibfnamefont{C.}~\bibnamefont{Gagliardi}},
  \bibinfo{author}{\bibfnamefont{V.}~\bibnamefont{Kroha}},
  \bibinfo{author}{\bibfnamefont{A.}~\bibnamefont{Mukhamedzanov}},
  \bibinfo{author}{\bibfnamefont{F.}~\bibnamefont{Nunes}},
  \bibinfo{author}{\bibfnamefont{X.}~\bibnamefont{Tang}},
  \bibinfo{author}{\bibfnamefont{L.}~\bibnamefont{Trache}}, \bibnamefont{and}
  \bibinfo{author}{\bibfnamefont{R.}~\bibnamefont{Tribble}},
  \bibinfo{journal}{Phys. Rev. C} \textbf{\bibinfo{volume}{63}},
  \bibinfo{pages}{055803} (\bibinfo{year}{2001}).

\bibitem[{\citenamefont{Tang et~al.}(2003)\citenamefont{Tang, Azhari,
  Gagliardi, Mukhamedzhanov, Pirlepesov, Trache, Tribble, Burjan, Kroha, and
  Carstoiu}}]{tang-prc67}
\bibinfo{author}{\bibfnamefont{X.}~\bibnamefont{Tang}},
  \bibinfo{author}{\bibfnamefont{A.}~\bibnamefont{Azhari}},
  \bibinfo{author}{\bibfnamefont{C.}~\bibnamefont{Gagliardi}},
  \bibinfo{author}{\bibfnamefont{A.}~\bibnamefont{Mukhamedzhanov}},
  \bibinfo{author}{\bibfnamefont{F.}~\bibnamefont{Pirlepesov}},
  \bibinfo{author}{\bibfnamefont{L.}~\bibnamefont{Trache}},
  \bibinfo{author}{\bibfnamefont{R.}~\bibnamefont{Tribble}},
  \bibinfo{author}{\bibfnamefont{V.}~\bibnamefont{Burjan}},
  \bibinfo{author}{\bibfnamefont{V.}~\bibnamefont{Kroha}}, \bibnamefont{and}
  \bibinfo{author}{\bibfnamefont{F.}~\bibnamefont{Carstoiu}},
  \bibinfo{journal}{Phys. Rev. C} \textbf{\bibinfo{volume}{67}},
  \bibinfo{pages}{015804} (\bibinfo{year}{2003}).

\bibitem[{\citenamefont{Ziegler et~al.}(1985)\citenamefont{Ziegler, Biersack,
  and Littmark}}]{Srim99}
\bibinfo{author}{\bibfnamefont{J.~F.} \bibnamefont{Ziegler}},
  \bibinfo{author}{\bibfnamefont{J.~P.} \bibnamefont{Biersack}},
  \bibnamefont{and} \bibinfo{author}{\bibfnamefont{U.}~\bibnamefont{Littmark}},
  \emph{\bibinfo{title}{The Stopping and Ranges of Ions in Matter, Vol I: The
  Stopping and Range of Ions in Solids}} (\bibinfo{publisher}{Pergamon Press,
  New York}, \bibinfo{year}{1985}).

\bibitem[{\citenamefont{Trache et~al.}(2003)\citenamefont{Trache, Azhari,
  Carstoiu, Clark, Gagliardi, Lui, Mukhamedzhanov, Tang, Timofeyuk, and
  Tribble}}]{trache-prc67}
\bibinfo{author}{\bibfnamefont{L.}~\bibnamefont{Trache}},
  \bibinfo{author}{\bibfnamefont{A.}~\bibnamefont{Azhari}},
  \bibinfo{author}{\bibfnamefont{F.}~\bibnamefont{Carstoiu}},
  \bibinfo{author}{\bibfnamefont{H.~L.} \bibnamefont{Clark}},
  \bibinfo{author}{\bibfnamefont{C.~A.} \bibnamefont{Gagliardi}},
  \bibinfo{author}{\bibfnamefont{Y.-W.} \bibnamefont{Lui}},
  \bibinfo{author}{\bibfnamefont{A.~M.} \bibnamefont{Mukhamedzhanov}},
  \bibinfo{author}{\bibfnamefont{X.}~\bibnamefont{Tang}},
  \bibinfo{author}{\bibfnamefont{N.}~\bibnamefont{Timofeyuk}},
  \bibnamefont{and} \bibinfo{author}{\bibfnamefont{R.~E.}
  \bibnamefont{Tribble}}, \bibinfo{journal}{Phys. Rev. C}
  \textbf{\bibinfo{volume}{67}}, \bibinfo{pages}{062801(R)}
  (\bibinfo{year}{2003}).

\bibitem[{\citenamefont{Trache et~al.}(2000)\citenamefont{Trache, Azhari,
  Clark, Gagliardi, Lui, Mukhamedzhanov, Tribble, and Carstoiu}}]{trache-prc61}
\bibinfo{author}{\bibfnamefont{L.}~\bibnamefont{Trache}},
  \bibinfo{author}{\bibfnamefont{A.}~\bibnamefont{Azhari}},
  \bibinfo{author}{\bibfnamefont{H.~L.} \bibnamefont{Clark}},
  \bibinfo{author}{\bibfnamefont{C.~A.} \bibnamefont{Gagliardi}},
  \bibinfo{author}{\bibfnamefont{Y.-W.} \bibnamefont{Lui}},
  \bibinfo{author}{\bibfnamefont{A.~M.} \bibnamefont{Mukhamedzhanov}},
  \bibinfo{author}{\bibfnamefont{R.~E.} \bibnamefont{Tribble}},
  \bibnamefont{and} \bibinfo{author}{\bibfnamefont{F.}~\bibnamefont{Carstoiu}},
  \bibinfo{journal}{Phys. Rev. C} \textbf{\bibinfo{volume}{61}},
  \bibinfo{pages}{024612} (\bibinfo{year}{2000}).

\bibitem[{\citenamefont{Jeukenne et~al.}(1977)\citenamefont{Jeukenne, Lejeune,
  and Mahaux}}]{JLM}
\bibinfo{author}{\bibfnamefont{J.~P.} \bibnamefont{Jeukenne}},
  \bibinfo{author}{\bibfnamefont{A.}~\bibnamefont{Lejeune}}, \bibnamefont{and}
  \bibinfo{author}{\bibfnamefont{C.}~\bibnamefont{Mahaux}},
  \bibinfo{journal}{Phys. Rev. C} \textbf{\bibinfo{volume}{16}},
  \bibinfo{pages}{80} (\bibinfo{year}{1977}).

\bibitem[{\citenamefont{Tang et~al.}(2004)\citenamefont{Tang, Azhari, Fu,
  Gagliardi, Mukhamedzhanov, Pirlepesov, Trache, Tribble, Burjan, Kroha
  et~al.}}]{tang-prc69}
\bibinfo{author}{\bibfnamefont{X.}~\bibnamefont{Tang}},
  \bibinfo{author}{\bibfnamefont{A.}~\bibnamefont{Azhari}},
  \bibinfo{author}{\bibfnamefont{C.}~\bibnamefont{Fu}},
  \bibinfo{author}{\bibfnamefont{C.~A.} \bibnamefont{Gagliardi}},
  \bibinfo{author}{\bibfnamefont{A.~M.} \bibnamefont{Mukhamedzhanov}},
  \bibinfo{author}{\bibfnamefont{F.}~\bibnamefont{Pirlepesov}},
  \bibinfo{author}{\bibfnamefont{L.}~\bibnamefont{Trache}},
  \bibinfo{author}{\bibfnamefont{R.~E.} \bibnamefont{Tribble}},
  \bibinfo{author}{\bibfnamefont{V.}~\bibnamefont{Burjan}},
  \bibinfo{author}{\bibfnamefont{V.}~\bibnamefont{Kroha}},
  \bibnamefont{et~al.}, \bibinfo{journal}{Phys. Rev. C}
  \textbf{\bibinfo{volume}{69}}, \bibinfo{pages}{055807}
  (\bibinfo{year}{2004}).

\bibitem[{\citenamefont{Blackmon et~al.}(2004)\citenamefont{Blackmon, Bardayan,
  Brune, Carstoiu, Champagne, Crespo, Davinson, Fernandes, Gagliardi, and
  et~al.}}]{blackmon-npa746}
\bibinfo{author}{\bibfnamefont{J.}~\bibnamefont{Blackmon}},
  \bibinfo{author}{\bibfnamefont{D.}~\bibnamefont{Bardayan}},
  \bibinfo{author}{\bibfnamefont{C.}~\bibnamefont{Brune}},
  \bibinfo{author}{\bibfnamefont{F.}~\bibnamefont{Carstoiu}},
  \bibinfo{author}{\bibfnamefont{A.}~\bibnamefont{Champagne}},
  \bibinfo{author}{\bibfnamefont{R.}~\bibnamefont{Crespo}},
  \bibinfo{author}{\bibfnamefont{T.}~\bibnamefont{Davinson}},
  \bibinfo{author}{\bibfnamefont{J.}~\bibnamefont{Fernandes}},
  \bibinfo{author}{\bibfnamefont{C.}~\bibnamefont{Gagliardi}},
  \bibnamefont{and} \bibinfo{author}{\bibfnamefont{U.~G.}
  \bibnamefont{et~al.}}, \bibinfo{journal}{Nucl. Phys. A}
  \textbf{\bibinfo{volume}{746}}, \bibinfo{pages}{365} (\bibinfo{year}{2004}).

\bibitem[{\citenamefont{{F. Carstoiu and L. Trache and R.~E.~Tribble and
  C.~A.~Gagliardi}}(2004)}]{carstoiu-prc70}
\bibinfo{author}{\bibnamefont{{F. Carstoiu and L. Trache and R.~E.~Tribble and
  C.~A.~Gagliardi}}}, \bibinfo{journal}{Phys. Rev. C}
  \textbf{\bibinfo{volume}{70}}, \bibinfo{pages}{054610}
  (\bibinfo{year}{2004}).

\bibitem[{\citenamefont{Trache et~al.}(1998)\citenamefont{Trache, Azhari,
  Clark, Gagliardi, Lui, Mukhamedzhanov, Tribble, and Carstoiu}}]{trache-prc58}
\bibinfo{author}{\bibfnamefont{L.}~\bibnamefont{Trache}},
  \bibinfo{author}{\bibfnamefont{A.}~\bibnamefont{Azhari}},
  \bibinfo{author}{\bibfnamefont{H.~L.} \bibnamefont{Clark}},
  \bibinfo{author}{\bibfnamefont{C.~A.} \bibnamefont{Gagliardi}},
  \bibinfo{author}{\bibfnamefont{Y.~W.} \bibnamefont{Lui}},
  \bibinfo{author}{\bibfnamefont{A.~M.} \bibnamefont{Mukhamedzhanov}},
  \bibinfo{author}{\bibfnamefont{R.~E.} \bibnamefont{Tribble}},
  \bibnamefont{and} \bibinfo{author}{\bibfnamefont{F.}~\bibnamefont{Carstoiu}},
  \bibinfo{journal}{Phys. Rev. C} \textbf{\bibinfo{volume}{58}},
  \bibinfo{pages}{2715} (\bibinfo{year}{1998}).

\bibitem[{\citenamefont{Bem et~al.}(2000)\citenamefont{Bem, Burjan, Kroha,
  Novak, Piskor, Simeckova, Vincour, Gagliardi, Mukhamedzhanov, and
  Tribble}}]{bem-prc62}
\bibinfo{author}{\bibfnamefont{P.}~\bibnamefont{Bem}},
  \bibinfo{author}{\bibfnamefont{V.}~\bibnamefont{Burjan}},
  \bibinfo{author}{\bibfnamefont{V.}~\bibnamefont{Kroha}},
  \bibinfo{author}{\bibfnamefont{J.}~\bibnamefont{Novak}},
  \bibinfo{author}{\bibfnamefont{S.}~\bibnamefont{Piskor}},
  \bibinfo{author}{\bibfnamefont{E.}~\bibnamefont{Simeckova}},
  \bibinfo{author}{\bibfnamefont{J.}~\bibnamefont{Vincour}},
  \bibinfo{author}{\bibfnamefont{C.~A.} \bibnamefont{Gagliardi}},
  \bibinfo{author}{\bibfnamefont{A.~M.} \bibnamefont{Mukhamedzhanov}},
  \bibnamefont{and} \bibinfo{author}{\bibfnamefont{R.~E.}
  \bibnamefont{Tribble}}, \bibinfo{journal}{Phys. Rev. C}
  \textbf{\bibinfo{volume}{62}}, \bibinfo{pages}{024320}
  (\bibinfo{year}{2000}).

\bibitem[{\citenamefont{Rhoades-Brown et~al.}(1980)\citenamefont{Rhoades-Brown,
  Mcfarlane, and Pieper}}]{PTOLEMY}
\bibinfo{author}{\bibfnamefont{M.}~\bibnamefont{Rhoades-Brown}},
  \bibinfo{author}{\bibfnamefont{M.~H.} \bibnamefont{Mcfarlane}},
  \bibnamefont{and} \bibinfo{author}{\bibfnamefont{S.~C.}
  \bibnamefont{Pieper}}, \bibinfo{journal}{Phys. Rev. C}
  \textbf{\bibinfo{volume}{21}}, \bibinfo{pages}{2417} (\bibinfo{year}{1980}).

\bibitem[{\citenamefont{Xu et~al.}(1994)\citenamefont{Xu, Gagliardi, Tribble,
  Mukhamedzhanov, and Timofeyuk}}]{hmxu-prl94}
\bibinfo{author}{\bibfnamefont{H.~M.} \bibnamefont{Xu}},
  \bibinfo{author}{\bibfnamefont{C.~A.} \bibnamefont{Gagliardi}},
  \bibinfo{author}{\bibfnamefont{R.~E.} \bibnamefont{Tribble}},
  \bibinfo{author}{\bibfnamefont{A.~M.} \bibnamefont{Mukhamedzhanov}},
  \bibnamefont{and} \bibinfo{author}{\bibfnamefont{N.~K.}
  \bibnamefont{Timofeyuk}}, \bibinfo{journal}{Phys. Rev. Lett.}
  \textbf{\bibinfo{volume}{73}}, \bibinfo{pages}{2027} (\bibinfo{year}{1994}).

\end{thebibliography}

\begin{widetext}
\begin{table*}
\caption{\label{tab_par} The optical model parameters, the corresponding $\chi^2$ per degree of freedom for
the $^7$Be elastic scattering fits, the calculated DWBA cross section for the $^{14}$N($^7$Be,$^8$B)$^{13}$C reaction (see text), and the ratio of the calculated DWBA calculation to that in Ref.\@ \protect\cite{azhari-prc60}.}
\begin{tabular}{cccccrcc}
\hline
Calculation & ~~$N_V$~~ & ~~$N_W$~~ & ~$t_V$~[fm]~ & ~$t_W$~[fm]~ & ~~$\chi^2/N$ &
~~~$\tilde\sigma_{DW}$~~~ & ~~Ratio~~ \\
 & & & & & & & 
${\cal R}=\frac{\sigma_{DW}({\small new})}{\sigma_{DW}({\small orig})}$ \\
\hline
Ref.\@ \cite{azhari-prc60} & 0.37 & 1.00 & 1.20 & 1.75 & 35.19 & 2.469 &1.000 \\
A & 0.45 & 0.90 & 1.20 & 1.75 & 10.72 &  & \\
B & 0.40 & 0.92 & 1.20 & 1.75 & 15.02 & 2.385 & 0.966\\
C & 0.42 & 0.92 & 1.20 & 1.75 & 12.84 &  & \\
D & 0.42 & 0.90 & 0.80 & 1.75 & 10.31 & 2.408 & 0.975\\
E & 0.42 & 0.90 & 0.80 & 1.55 &  7.97 & 2.538 & 1.028\\
F & 0.52 & 0.78 & 0.12 & 2.59 & 25.72 &  & \\
G & 0.40 & 0.85 & 1.20 & 1.75 &  7.49 & 2.375 & 0.962 \\
H & 0.40 & 0.85 & 1.20 & 1.75 &  9.22 & 2.137 & 0.900\\
\hline
\end{tabular}
\end{table*}
\end{widetext}

\end{document}